\documentclass{article}

\usepackage{arxiv}

\usepackage[utf8]{inputenc} 
\usepackage[T1]{fontenc}    
\usepackage[bookmarks=false]{hyperref}       
\usepackage{url}            
\usepackage{booktabs}       
\usepackage{amsfonts}       
\usepackage{nicefrac}       
\usepackage{microtype}      
\usepackage{lipsum}		
\usepackage{graphicx}

\title{COBI-GRINE: A Tool for Visualization and Advanced Evaluation of Communities in Mass Channel Similarity Graphs}

\author{\small \href{https://orcid.org/0000-0002-3935-7598}{\includegraphics[scale=0.06]{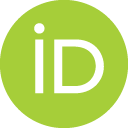}\hspace{1mm}Karsten Wüllems} \\
	\small International DFG Research Training Group GRK 1906\\
	\small Biodata Mining Group, Faculty of Technology\\	
	\small Center for Biotechnology (CeBiTec)\\
	\small Bielefeld University\\
	\small Bielefeld, Germany \\
	\small \texttt{wuellems@cebitec.uni-bielefeld.de} \\
	\And
	\small Daniel Göbel \\
	\small Biodata Mining Group, Faculty of Technology\\	
	\small Bielefeld University\\
	\small Bielefeld, Germany \\
	\And
	\small Annika Zurowietz \\
	\small Proteome and Metabolome Research, Faculty of Biology\\	
	\small Bielefeld University\\
	\small Bielefeld, Germany \\
	\And
	\small Hanna Bednarz \\
	\small Proteome and Metabolome Research, Faculty of Biology\\	
	\small Bielefeld University\\
	\small Bielefeld, Germany \\
	\And
	\small Karsten Niehaus \\
	\small Proteome and Metabolome Research, Faculty of Biology\\	
	\small Bielefeld University\\
	\small Bielefeld, Germany \\
	\And
	\small Tim W. Nattkemper \\
	\small Biodata Mining Group, Faculty of Technology\\	
	\small Center for Biotechnology (CeBiTec)\\
	\small Bielefeld University\\
	\small Bielefeld, Germany \\
	\small \texttt{tim.nattkemper@uni-bielefeld.de} \\
}

\renewcommand{\shorttitle}{Grine}

\hypersetup{
pdftitle={COBI-GRINE: A Tool for Visualization and Advanced Evaluation of Communities in Mass Channel Similarity Graphs},
pdfsubject={stat.CO},
pdfauthor={Karsten Wüllems, Daniel Göbel, Annika Zurowietz, Hanna Bednarz, Karsten Niehaus, Tim W. Nattkemper},
pdfkeywords={MALDI imaging, Networks, Clustering, Community detection, Visualization, Graphs},
}

\begin{document}
\maketitle

\begin{abstract}
The detection of groups of molecules that co-localize with histopathological patterns or sub-structures is an important step to combine the rich high-dimensional content of mass spectrometry imaging (MSI) with classic histopathological staining.
Here we present the evolution of GRINE to COBI-GRINE, an interactive web tool that maps MSI data onto a graph structure to detect communities of laterally similar distributed molecules and co-visualizes the communities with Hematoxylin and Eosin (HE) stained images. Thereby the tool enables biologists and pathologists to examine the MSI image graph in a target-oriented manner and links molecular co-localization to pathology. Another feature is the manual optimization of cluster results with the assist of graph statistics in order to improve the community results.
As the graphs can become very complex, those statistics provide good heuristics to support and accelerate the detection of sub-clusters and misclusterings.
This kind of edited cluster optimization allows the integration of expert background knowledge into the clustering result and a more precise analysis of links between molecular co-localization and pathology.
\end{abstract}


\newpage
\section{Introduction}
Cluster analysis is one of the most common unsupervised analysis method used in computer aided exploration of mass spectrometry imaging (MSI) data.
There are two main approaches to cluster MSI data: 1. clustering pixels according to mass spectra similarities or 2. clustering of mass channels according to similarities in molecular lateral distribution patterns. The result is usually displayed as a segmentation map. The second approach groups mass channels according to their co-localization, expressed by similar distributions. As mass channels usually do not co-localize perfectly and a definition of a distance function for lateral distributions is not straight forward, this clustering result is a little more ambiguous than segmentation maps. While many works about the pixel clustering approach are available \cite{kolling2012whide}\cite{alexandrov2010spatial}\cite{deininger2008maldi} the works published for mass channel clustering are less prominent, with GRINE \cite{wullems2019detection} as one example.
As the co-localization of mass channels often appears fuzzy and ambiguous, we propose to explore the cluster results in combination with other image modalities, such as optical scans or histologically stained consecutive slides.
Here we present the visual exploration tool COBI-GRINE (analysis, \textbf{c}luster \textbf{o}ptimization and \textbf{b}iological \textbf{i}nterpretation of \textbf{gr}aph mapped \textbf{i}mage data \textbf{ne}tworks). COBI-GRINE features new tools to visually explore but also edit co-localizing mass channel communities. One main feature is the option to link and fuse the community visualizations with other modalities, like Hematoxylin and Eosin (HE) stained sections, which paves the way to link groups of co-localizing molecules to distinct tissue substructures. To the best of our knowledge, there is no MSI visualization tool that features these functions.

\section{Experimental Section}
COBI-GRINE is a webtool, developed for the use in combination with a localhost. The backend is written in Python using Flask as web application framework. The frontend is written in JavaScript using Vue as JavaScript framework and BootstrapVue as CSS framework. To facilitate its use, COBI-GRINE runs in a Docker container. This way the user only needs Docker and a web browser (Chrome or Firefox are recommended). The source code is available under the GNU GPLv3 license at \url{https://github.com/Kawue/grine-v2/}.

\section{Results and Discussion}
\begin{figure}[h]
    \centering
    \includegraphics[width=1\linewidth]{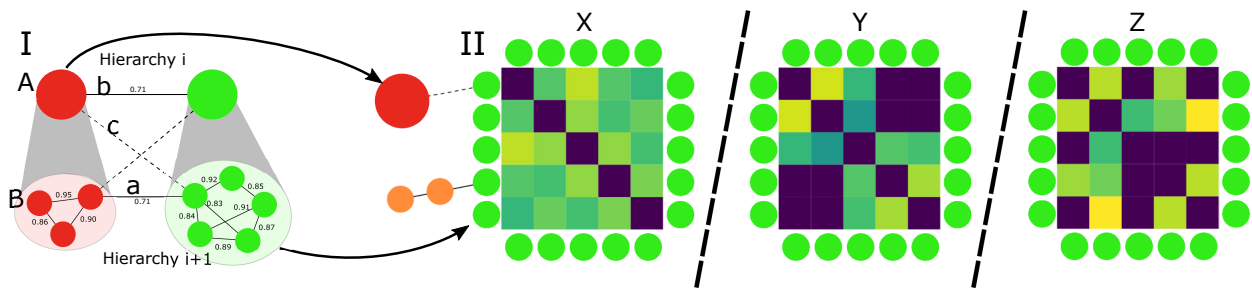}
    \caption{I. Outline of the Mass Channel Similarity Graph (MCSG) structure. Nodes can either be mass channel nodes (A) or community nodes (B). Edges can be mass channel edges (a) if they connect two mass channel nodes, community edges (b) if they connect two communities or hybrid edges (c) if a mass channel node is connected with another mass channel node hidden within a community. The weight of mass channel edges correspond to the similarity between the respective mass channel images, while the weight of community edges corresponds to the mean weight of all edges that run between mass channel nodes of the respective communities. II. The NodeTrix presentation with examples for a cluster that is homogeneous (X), builds subclusters (Y) or is heterogeneous (Z).}
    \label{fig:graphstructure-nodetrix}
\end{figure}

COBI-GRINE visualizes hierarchically organized communities (or clusters) computed from so called Mass Channel Similarity Graph (MCSG). A MCSG represents the strengths of the molecular co-localizations in one MSI data set. It contains two types of nodes: 1. mass channel nodes, representing a single mass channel image and 2. community nodes, representing a cluster of mass channel images. Edges can be classified into three categories: 1. mass channel edges, connecting two mass channel nodes, with a length proportional to the co-localization of the respective mass channels, 2. community edges, connecting two different communities, with a length proportional to the mean co-localization of all mass channel edges that run between both communities and 3. hybrid edges, connecting a mass channel node and a community node, which indicates a connection to a mass channel node hidden within the community node. The weight of hybrid edges is a very small constant since they are only visual indicators and should not affect the graph layout algorithm. The whole MCSG structure is illustrated in Figure~\ref{fig:graphstructure-nodetrix}~I.

A script presented in our previous work \cite{wullems2019detection} performs both, graph mapping and community detection, and provides all necessary files to run COBI-GRINE. The respective code is available at \url{https://github.com/Kawue/msi-community-detection} under the under the GNU GPLv3 license.

For a detailed explanation of the basic interactions and functionalities we refer to our previous work on GRINE \cite{wullems2019detection}. However, the visual design of the tool has been changed significantly to integrate the new functions and for the sake of a better usability. An example screenshot overview is shown in Figure~\ref{fig:interface}. In summary, COBI-GRINE offers the following four main innovations to make the community detection on MCSG's more accessible and to increase its capabilities for biologists and pathologists:

\begin{figure}[h]
    \centering
    \includegraphics[width=0.8\linewidth]{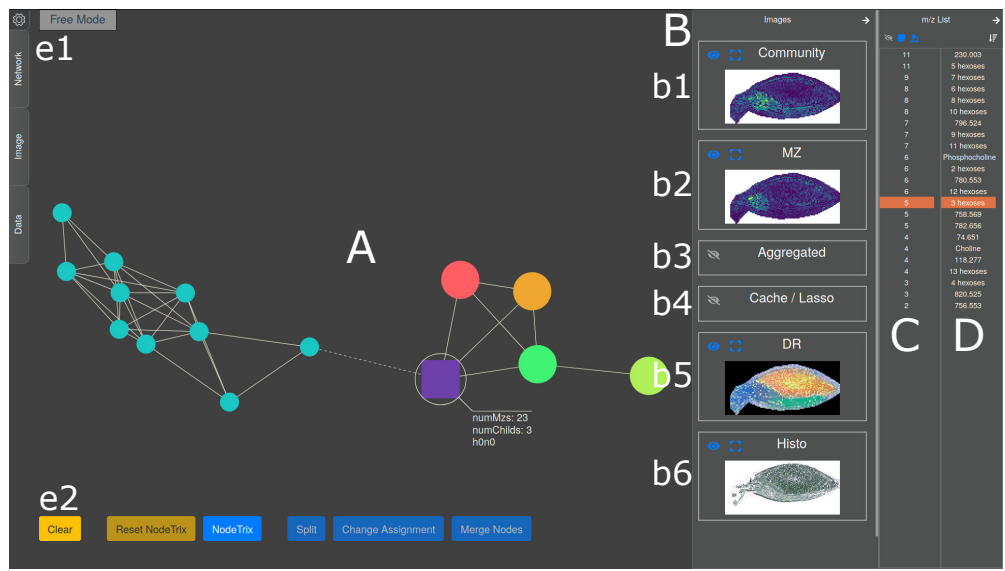}
    \caption{Overview of the COBI-GRINE interface. (A) shows the graph panel with an expanded community and an active hover annotation on the selected community. (B) shows the image panel, where all images are displayed. (b1) - (b4) are different mass channel images or mass channel image stacks. (b5) is a RGB three component dimension reduction projection as described in our previous work \cite{wullems2019detection}. (b6) shows an optical image. The lasso region selection is available for (b4) and (b6). (C) and (D) are lists that display QGP values and mass channel values (or annotations), respectively. (e1) and (e2) are settings elements.}
    \label{fig:interface}
\end{figure}

\noindent \textbf{NodeTrix Representation:} NodeTrix \cite{henry2007nodetrix} represents the adjacency matrix of a selected subset of graph nodes as a heatmap, visually embedded in the entire graph structure. This is useful to visually explore densely connected subgraphs and to provide a very fast overview about the structural features of the selected subset of nodes, i.e.\ observation of overall strong interconnection, overall weak interconnection or strongly interconnected subgroups. Thus the heatmap can reveal if a subset is homogeneous, heterogeneous or builds subclusters (exemplified in Figure~\ref{fig:graphstructure-nodetrix}~II). This can be of great benefit to assess the quality of a cluster or to detect starting points for cluster modifications.

\begin{figure}
    \centering
    \includegraphics[width=0.8\linewidth]{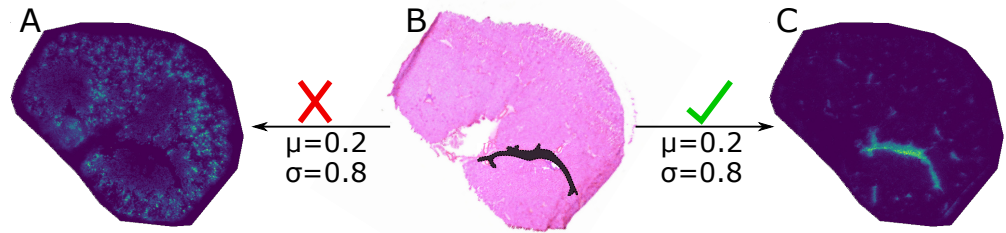}
    \caption{Outline of the image guided lasso exploration. a}
    \label{fig:image-guided-exploration}
\end{figure}

\noindent \textbf{Image Guided Exploration:} If any aligned image modality is provided, a lasso selection on this modality can be used to select a region of interest. COBI-GRINE will visually indicate all nodes that correspond to mass channel images with at least $\sigma$ pixels of minimum intensity $\mu$ within the selected region. Both parameters $\mu$ and $\sigma$ are user defined percentage thresholds for values (exemplified in Figure~\ref{fig:image-guided-exploration}).

\noindent \textbf{Manual Cluster Modification:} As mentioned before the clustering of mass channel images leads to results that require posterior edits by the user. A new intuitive interface allows to merge and split communities or to change the community assignment of single nodes. The modified MCSG can be exported as JSON file and used by other researchers. This way the results can be easily stored and shared.

\noindent \textbf{Quantitative Graph Property Guided Exploration:} To further use the advantages of the graph structure we have adopted a proposal from our previous work and implemented a set of quantitative graph properties (QGP). These QGP's can be used as a heuristic to identify nodes with special characteristics, like community cores (hubs), potential singletons, wrong assignments and bridges. For a large and complex graph, QGS's can serve as a starting point for further detailed analysis.

\section{Conclusion}
COBI-GRINE is a comprehensive extension of the previously presented GRINE framework. The new features simplify the exploration and evaluation of MCSG communities, they enable the integration of expert knowledge into the community detection result by manual modification and they extend the field of application to areas such as digital histopathology by integrating other image modalities.

\section*{Availability}
Source code is available on github.\\
\textbf{Project name:} Grine-v2 \\
\textbf{Project home page:} \href{https://github.com/Kawue/grine-v2/}{https://github.com/Kawue/grine-v2/} \\
\textbf{Operating system(s):} Platform independent \\
\textbf{Programming language:} Python \\
\textbf{Other requirements:} Python 3.5 or higher \\
\textbf{License:} GNU GPLv3 \\

\section*{Acknowledgement}
We acknowledge the financial support of KW by the German Research Foundation (DFG) as part of the GRK 1906.

\bibliographystyle{unsrt}
\bibliography{ms}  

\end{document}